\documentclass[a4paper]{spie}  
\usepackage[pdftex]{graphicx}
\usepackage[amssymb]{SIunits}
\usepackage{amssymb}
\usepackage{amsmath}

\title{Generation of a two-photon state from a quantum dot in a
  microcavity under incoherent and coherent continuous excitation}

\author{Elena del Valle,\supit{a} Alejandro Gonzalez-Tudela,\supit{b} and Fabrice P. Laussy\supit{c}
\skiplinehalf
\begin{small}
  \supit{a} Physikdepartment,  Technische Universit\"at M\"unchen, James-Franck-Str. 1, 85748 Garching, Germany; \\
  \supit{b} F\'isica Te\'orica de la Materia Condensada, Universidad Aut\'onoma de Madrid, 28049, Madrid, Spain; \\
  \supit{c} Walter Schottky Institut, Technische Universit\"at
  M\"unchen, Am Coulombwall 3, 85748 Garching, Germany.
\end{small}
}

\authorinfo{Further author information: (Send correspondence to E. del Valle)\\E. del Valle: E-mail: elena.delvalle.reboul@gmail.com}

\begin{document} 
  \maketitle 
\newcommand{\ud}[1]{#1^{\dagger}} 
\newcommand{\bra}[1]{\left\langle #1\right|}
\newcommand{\ket}[1]{\left| #1\right\rangle}
\newcommand{\braket}[2]{\langle #1|#2\rangle}
\newcommand{\Imamoglu}{\u Imamo\=glu}
\newcommand{\mean}[1]{\langle#1\rangle}

\begin{abstract}
  We analyze the impact of both an incoherent and a coherent
  continuous excitation in our proposal to generate a two-photon state
  from a quantum dot in a microcavity [New J.~Phys.~\textbf{13},
  113014 (2011)]. A comparison between exact numerical results and
  analytical formulas provides the conditions to efficiently generate
  indistinguishable and simultaneous pairs of photons under both types
  of excitation.
\end{abstract}




\section{INTRODUCTION}
\label{sec:intro}  

A single quantum dot in a cavity can be turned into a two-photon
emitter by tuning the cavity frequency into resonance with half the
biexciton energy~\cite{delvalle10a}. Since the biexciton frequency can
be far from twice the exciton energy---thanks to the binding
energy---the exciton states can be very far from the cavity too. This
allows their emission through the cavity to be suppressed while
Purcell-enhancing the two-photon emission from the
biexciton~\cite{delvalle11d}. The principle has been recently realised
in a system well into the strong coupling regime~\cite{ota11a}, where
the authors have observed a strong enhancement of the
photoluminescence spectrum at the two-photon resonance (2PR),
compatible with very low number of photons. This means that the
emission was in the spontaneous (or linear) regime. In this
experiment, the system was excited via a continuous excitation of the
wetting layer, resulting in an incoherent population of the dot
levels. 

\emph{Incoherent excitation} is a convenient way of probing the
system, typically used to characterised the level structure and
optical resonances. However, it may have a substantial and undesired
impact on the quantum properties, namely, an increase of decoherence
and dephasing in the dynamics (even more so when the features of
interest require a pumping to high rungs of excitation, as it is the
case for the Jaynes-Cummings nonlinearities~\cite{arXiv_laussy12b},
one atom laser~\cite{gartner11a,delvalle10b,delvalle11a} or entangled
photon pair generation~\cite{arXiv_poddubny12a}, among others). In
general, it is advisable to include incoherent excitation when
describing experimental results in order to interpret them correctly
(e.~g.~strong coupling~\cite{laussy08a,delvalle10b,poddubny10a},
superradiance~\cite{auffeves11a,averkiev09a}, phase
transitions~\cite{arXiv_laussy12a}, etc.)

\emph{Coherent excitation}, close to resonance to the quantum dot
levels, provides a second possibility to probe the system. To this
intent, one can for instance apply a laser whose polarization is
orthogonal to that of the cavity mode, in order not to excite cavity
photons (directly or indirectly though the state with the cavity
polarization). Coherent excitation may seem like a better choice
because it does not introduce extra decoherence in the
dynamics. Moreover, if the laser is also tuned to the two-photon
resonance, it will excite the biexciton directly with high
probability~\cite{stufler06a}. However, one must remain in the linear
regime as well in order not to dress the system~\cite{muller08a},
adding excitation-induced features.

In this work, we study theoretically the two-photon emission under a
low continuous excitation of both types. In Section~\ref{sec:1}, we
obtain numerically the dot-cavity steady state and locate the one- and
two-photon resonances in the system. In Section~\ref{sec:2}, we
develop an analytical approach at the two-photon resonance, consisting
in solving the effective dot dynamics and deriving the cavity
properties from physical arguments. The comparison between the
numerical and analytical results provides the limiting excitation
before the two-photon emission is hindered by decoherence or dressed
states. The formulas provide optimum pumping strength and information
on the differences between the two types of excitation.

\section{TWO-PHOTON RESONANCE UNDER CONTINUOUS EXCITATION}
\label{sec:1}

\begin{figure}[t]
  \centering
  \includegraphics[width=.8\linewidth]{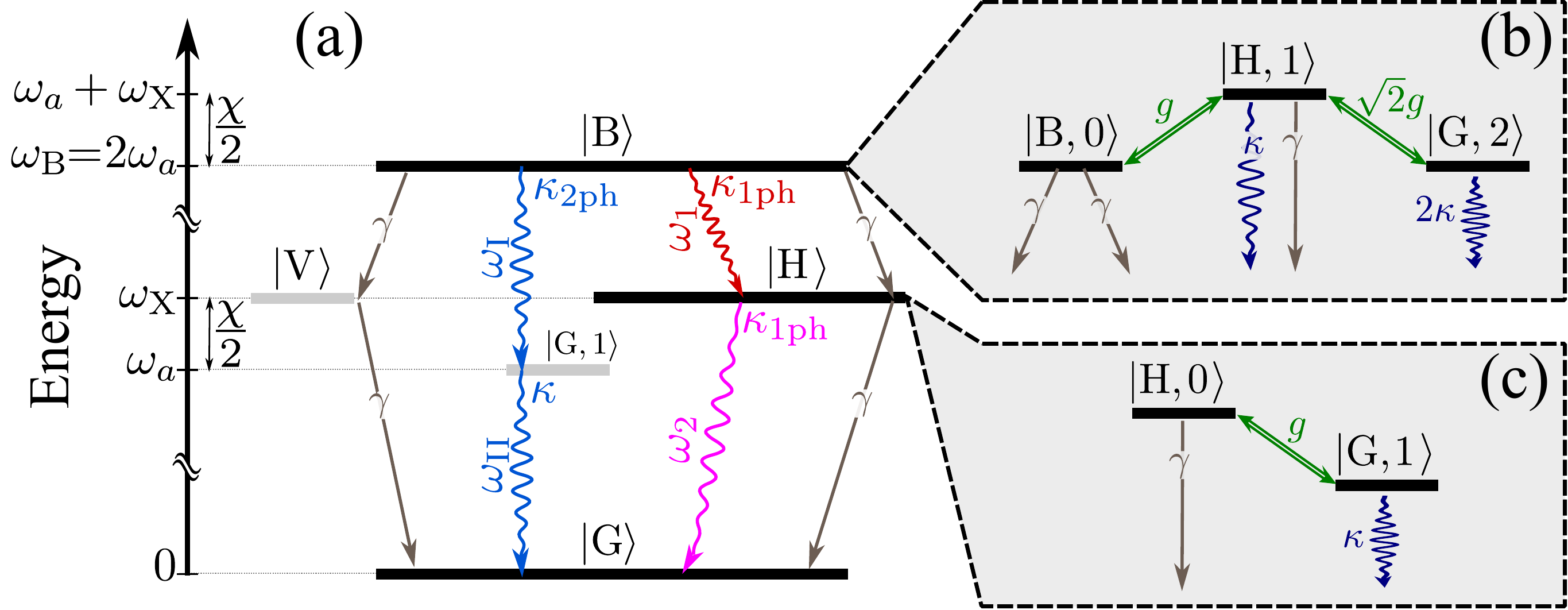}
  \caption{Level scheme of a quantum dot coupled to a cavity mode with
    linear polarization~H at the 2PR. In (a), the quantum dot
    four-level system is shown including all loss channels: on the one
    hand, the excitonic spontaneous decay, $\gamma$, and the
    one-cavity-photon decay, $\kappa_\mathrm{1P}$, taking place at
    frequencies $\omega_1$ and~$\omega_2$; on the other hand, the
    two-cavity-photon Purcell enhanced decay, $\kappa_\mathrm{2P}$, at
    the cavity frequency $\omega_\mathrm{I,II}\approx \omega_a$. In
    (b) is shown how (thanks to the dispersive coupling to the cavity)
    the state $\ket{\mathrm{B}}\approx \ket{\mathrm{B,0}}$ acquires a
    small component from other states with one and two photons, giving
    rise to the effective one- and two-photon decay rates
    $\kappa_\mathrm{1P}$ and $\kappa_\mathrm{2P}$. In (c), id.~but for
    the state $\ket{\mathrm{H}}\approx \ket{\mathrm{H,0}}$ that
    acquires a small one-photon component that leads to
    $\kappa_\mathrm{1P}$.}
  \label{fig:1}
\end{figure}

The system consists of a dot embedded in a microcavity, coupled only
to one of its modes, polarised along a direction that we call
horizontal (H). It is, therefore, convenient to study the dot in the
basis of the two orthogonal linear polarizations H and V:
$\{\ket{\mathrm{G}},\ket{\mathrm{V}},\ket{\mathrm{H}},\ket{\mathrm{B}}\}$,
where G stands for the ground state, V for vertically polarised dot
state and B for the biexciton state, that is, having both spin-up and
spin-down states (or H and V polarised states) excited. The total
Hilbert space including the cavity degree of freedom is expanded in
terms of the basis
$\{\ket{\mathrm{G}n},\ket{\mathrm{V}n},\ket{\mathrm{H}n},\ket{\mathrm{B}n}\}$
with $n=0,\hdots,\infty$ is the number of photons. The Hamiltonian of
the system reads ($\hbar=1$):
\begin{equation}
  \label{eq:ThuApr14002810CEST2011}
  H_\mathrm{dot-cav}=\underbrace{\omega_\mathrm{X}\big(\ket{\mathrm{V}}\bra{\mathrm{V}}+\ket{\mathrm{H}}\bra{\mathrm{H}}\big)  +(2\omega_\mathrm{X}-\chi) \ket{\mathrm{B}}\bra{\mathrm{B}}}_\text{$H_\mathrm{dot}$}+\underbrace{\omega_a\ud{a}a}_\text{$H_\mathrm{cavity}$}+ \underbrace{g\sum_{i=\mathrm{H,V}}\big[\ud{a} (\ket{\mathrm{G}}\bra{i}+\ket{i}\bra{\mathrm{B}})+ \text{h. c.}\big]}_\text{$H_\mathrm{coupling}$} \,,
\end{equation}
where $a$ is the cavity field annihilation operator (boson) with
frequency~$\omega_a$. We consider the excitonic states at the same
frequency $\omega_\mathrm{X}$ since a possible splitting between them
does not affect our results. The binding energy of the biexciton
state, $\chi$, is the key parameter to turn the system into a
two-photon emitter. If the cavity is tuned so that two photons match
the biexciton frequency, $\omega_a=\omega_\mathrm{X}-\chi/2$, as in
Fig.~\ref{fig:1}(a), the paired emission of cavity photons is
enhanced by two-photon Purcell effect~\cite{delvalle10a,delvalle11d}.

Dissipation and excitation are included in the master equation
\begin{equation}
  \label{eq:ThuDec15142224CET2011}
  \partial_t\tilde \rho= i[\tilde\rho, H_\mathrm{dot-cav}]+\mathbf{\tilde L}_\mathrm{decay}(\tilde \rho)+\mathbf{L}_\mathrm{coh}(\tilde \rho)+\mathbf{L}_\mathrm{incoh}(\tilde \rho)\,,
\end{equation}
in the following form:
\begin{subequations}
  \label{eq:TueDec20144239CET2011}
  \begin{align}
    &\mathbf{\tilde L}_\mathrm{decay}(\tilde\rho)=\frac{\kappa}{2}\mathcal{L}_a(\tilde\rho)+\frac{\gamma}{2}\sum_{i=\mathrm{V,H}}\Big[\mathcal{L}_{\ket{\mathrm{G}}\bra{i}}+\mathcal{L}_{\ket{i}\bra{\mathrm{B}}}\Big](\tilde\rho)\,,\\
    &\mathbf{L}_\mathrm{coh}(\tilde \rho)=i\Omega_\mathrm{V} [\tilde
    \rho,\ket{\mathrm{G}}\bra{\mathrm{V}}+\ket{\mathrm{V}}\bra{\mathrm{B}}+\text{h. c.}]\,,\\
    &\mathbf{L}_\mathrm{incoh}(\tilde
    \rho)=\frac{P}{2}\sum_{i=\mathrm{V,H}}\Big[\mathcal{L}_{\ket{\mathrm{i}}\bra{\mathrm{G}}}+\mathcal{L}_{\ket{\mathrm{B}}\bra{\mathrm{i}}}\Big](\tilde
    \rho)\,.
  \end{align}
\end{subequations}
where~$\mathcal{L}_c(\rho)=2c\rho\ud{c}-\ud{c}c\rho-\rho \ud{c}c$ is
in the Lindblad form. We call $\kappa$ the cavity losses and $\gamma$
the exciton relaxation rates. The variable $\tilde \rho$ refers to the
total cavity-dot density matrix while $\rho$ refers to the reduced dot
density matrix, tracing out the cavity degree of freedom. We consider
the separate action of two types of continuous excitation: coherent
excitation (resonant driving of the dot) with vertical polarization,
$\mathbf{L}_\mathrm{coh}$, and incoherent excitation (off-resonant
driving of the wetting layer) affecting all four levels,
$\mathbf{L}_\mathrm{incoh}$. In the case of coherent excitation, the
master equation is in a frame rotating at the laser frequency and,
therefore, all frequencies are referred to the laser one~$\omega_L$,
which we set at the two-photon resonance to maximise the biexciton
population: $\omega_L=\omega_\mathrm{X}-\chi/2$.

We solve equation Eq.~(\ref{eq:ThuDec15142224CET2011}) in the steady
state by setting $\partial_t \tilde \rho=0$. First, this is done
numerically with a sufficient truncation in the number of photons. We
take the exciton relaxation rate, $\gamma$, as the smallest parameter
in the system and the biexciton binding energy, $\chi$, as the
largest, which is the typical experimental situation~\cite{ota11a}. In
order to increase efficiency of the two photon emission, the system is
in the regime of strong coupling, that is, we assume parameters in the
range~\cite{delvalle11d} $\gamma \ll \kappa\lessapprox g \ll \chi$.

\begin{figure}[t]
  \centering
  \includegraphics[width=\linewidth]{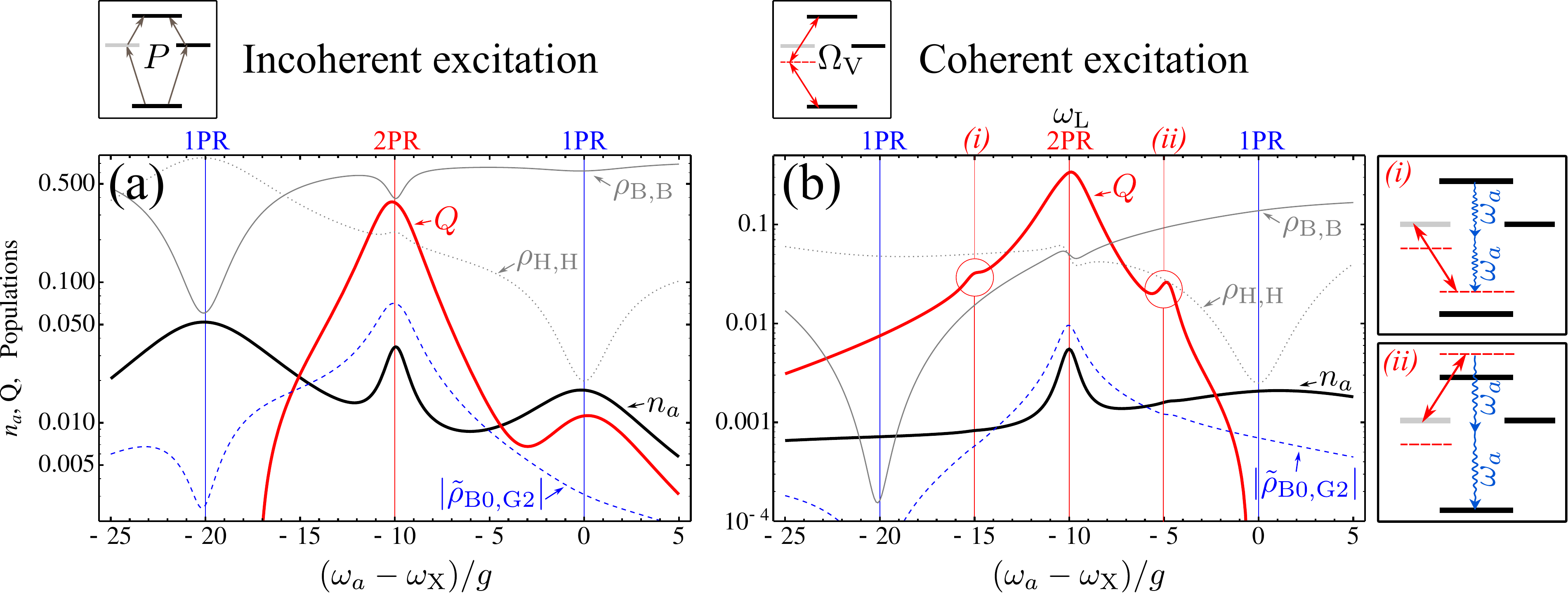}
  \caption{Steady state of the system under (a) incoherent and (b)
    coherent excitation when sweeping the cavity frequency through the
    different resonances. The results are exact, obtained by solving
    numerically the corresponding full master
    equation~(\ref{eq:ThuDec15142224CET2011}). We plot $Q$, $n_a$,
    $\rho_\mathrm{B,B}$, $\rho_\mathrm{H,H}$ and
    $|\tilde\rho_\mathrm{B0,G2}|$. With vertical guide lines we have
    marked the two-photon and one-photon resonances. The laser
    frequency in (b) is set at the two-photon resonant excitation,
    $\omega_\mathrm{L}=\omega_\mathrm{X}-\chi/2$. In this case, we
    also find two additional bunching peaks in $Q$, (i) and (ii), due
    to the two-cavity-photon resonance with virtual (Raman) states
    driven by the laser. Parameters: $\chi=20g$, $\kappa=g$,
    $\gamma=0.01g$, $P=0.06g$, $\Omega_\mathrm{V}=0.5g$.}
  \label{fig:2}
\end{figure}

The main quantities of interest---characterising the cavity
emission---are the steady state mean cavity photon number $n_a=\mean{
  a^\dagger a}$ and the $Q$ Mandel factor, $Q=n_a(G^{(2)}/n_a^2-1)$,
related to the second order coherence function at zero delay,
$G^{(2)}=\mean{a^\dagger a^\dagger aa}$. The $Q$ factor quantifies
bunching ($Q>0$) and antibunching ($Q<0$) in the emission, taking into
account the available signal $n_a$. We plot $Q$ and $n_a$ in
Fig.~\ref{fig:2} under weak incoherent (a) and coherent (b)
excitation.

As a first step, we tune the cavity frequency through the system,
probing the different resonances. Keeping in mind the level structure
of Fig.~\ref{fig:1}, we locate the two-photon resonance (2PR) by a
clear bunching peak of the $Q$-factor at $\omega_a \approx
\omega_\mathrm{B}/2= \omega_\mathrm{X}-\chi/2$ (at $-10g$ in the
figures). This corresponds to the simultaneous emission of two cavity
photons from $\ket{\mathrm{B}}$ as shown in
Refs.~\cite{delvalle10a,delvalle11d}. On the other hand, at the two
possible one-photon resonances (1PR), namely at
$\omega_a=\omega_1\approx \omega_\mathrm{X}-\chi$ and
$\omega_a=\omega_2\approx \omega_\mathrm{X}$ (at $-20g$ and $0$ in the
figures), the $Q$-factor drops or even becomes negative. These
features, accompanied by an enhancement in the cavity emission $n_a$,
especially at the 2PR, are in agreement with the properties found for
an ideal device that can be prepared in the biexciton state
$\ket{\mathrm{B}0}$~\cite{delvalle11d}. There are, however, some
differences between this ideal case and the two types of
excitation. For instance, under coherence excitation, two new
two-photon resonances appear at $\omega_\mathrm{X}-3\chi/4$ and
$\omega_\mathrm{X}-\chi/4$ (at $-15g$ and $-5g$ in the figure) that we
call $(i)$ and $(ii)$. They arise when two photons match the
transitions to Raman virtual states created by the laser, close to
$\ket{\mathrm{G}}$ and $\ket{\mathrm{B}}$, respectively, as depicted
on the right hand side of Fig.~\ref{fig:2}. These are excitation
induced resonances. They show how the cavity emission can be strongly
and qualitatively affected by the laser even though it has the
orthogonal polarization.

In order to analyse other more subtle issues related to the
excitation, regarding efficiency, degree of simultaneity or
indistinguishability of the two-photon emission at the 2PR, we will
carry out some approximations on Eq.~(\ref{eq:ThuDec15142224CET2011})
and obtain analytical expressions for the dot populations, $n_a$ and
$Q$ .

\section{ANALYTICAL RESULTS}
\label{sec:2}

Let us tune the cavity to the two-photon resonance (2PR),
$\omega_a\approx \omega_\mathrm{X}-\chi/2$, as in Fig.~\ref{fig:1}.
Here, the effective one-photon and two-photon coupling strengths are
much smaller than the cavity decay rate, $g_\mathrm{1P}\approx g
\kappa/\chi$ and $g_\mathrm{2P}\approx
4g^2/(\sqrt{2}\chi)\ll\kappa$~\cite{delvalle10a}. Therefore, we can
adiabatically eliminate the cavity and consider only the quantum dot
effective dynamics in its reduced Hilbert space. The cavity simply
provides three extra decay channels that are Purcell
suppressed/enhanced, given by the rates~\cite{delvalle11d}:
\begin{equation}
\label{eq:MonDec12101215CET2011}
\kappa_\mathrm{1P}= 4 g_\mathrm{1P}^2/\kappa\quad \text{and}\quad\kappa_\mathrm{2P}=4 g_\mathrm{2P}^2/(2\kappa)\,.
\end{equation}
The first one provides a second de-excitation channel from
$\ket{\mathrm{B}}$ to $\ket{\mathrm{H}}$ and from $\ket{\mathrm{H}}$
to $\ket{\mathrm{G}}$ via the emission of one cavity photon. The
second rate provides a third de-excitation channel from
$\ket{\mathrm{B}}$ to $\ket{\mathrm{G}}$ via the emission of two
cavity photons. 

\subsection{Quantum dot properties}

The master equation for the reduced dot density matrix~$\rho$, where
the cavity degree of freedom has been traced out, reads:
\begin{subequations}
  \label{eq:ThuApr174005647CEST2011}
  \begin{align}
    \partial_t\rho=&i[\rho,H_\mathrm{dot}]+\mathbf{L}_\mathrm{decay}(\rho)+\mathbf{L}_\mathrm{coh}(\rho)+\mathbf{L}_\mathrm{incoh}(\rho)\,,\\
    \mathbf{L}_\mathrm{decay}(\rho)=&\frac{\gamma+\kappa_{1P}}{2}\Big[\mathcal{L}_{\ket{\mathrm{G}}\bra{H}}+\mathcal{L}_{\ket{H}\bra{\mathrm{B}}}\Big](\rho)+\frac{\kappa_{2P}}{2}\mathcal{L}_{\ket{\mathrm{G}}\bra{B}}(\rho)+\frac{\gamma}{2}\Big[\mathcal{L}_{\ket{\mathrm{G}}\bra{V}}+\mathcal{L}_{\ket{V}\bra{\mathrm{B}}}\Big](\rho)
    \,.
  \end{align}
\end{subequations}
For simplicity, we neglect the small Stark shifts induced by the
dispersive coupling on the exciton and cavity frequencies, of the
order of~$g_\mathrm{1P}$, $g_\mathrm{2P}$. 

\begin{figure}[t]
  \centering
  \includegraphics[width=\linewidth]{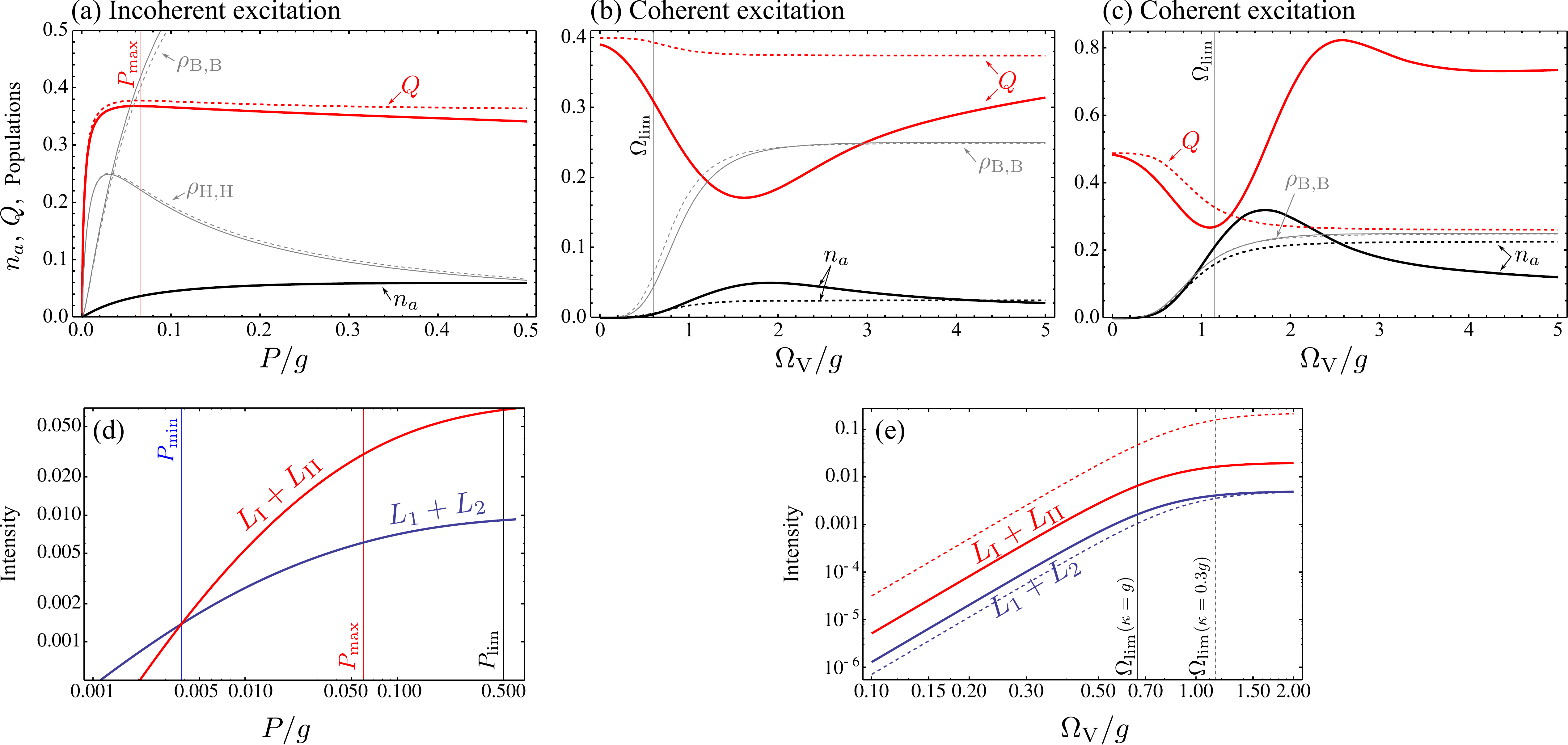}
  \caption{\newline \underline{First row}: Steady state of the system
    at the 2PR under (a) incoherent and (b), (c) coherent excitation
    as a function of the corresponding excitation rates. The
    quantities $Q$, $n_a$, $\rho_\mathrm{B,B}$ and $\rho_\mathrm{H,H}$
    are shown.  The exact numerical results appear with solid lines
    and the analytical approximations discussed in the text in dashed
    lines. The formulas for $n_a$ and $Q$ are a good approximation for
    $P<P_\mathrm{lim}$ and $\Omega_\mathrm{V}<\Omega_\mathrm{lim}$.
    The maximum $Q$ under incoherent excitation is achieved at
    $P_\mathrm{max}=0.066g$, which is the parameter used in
    Fig.~\ref{fig:2}(a). The optimum $Q$ under coherent excitation is
    achieved at vanishing pumping.  Parameters: $\chi=20g$,
    $\omega_a=\omega_\mathrm{X}-\chi/2$.  In (a), (b) $\kappa=g$,
    giving $\kappa_\mathrm{1P}=0.01g$, $\kappa_\mathrm{2P}=0.04g$,
    $P_\mathrm{lim}=0.5g$ and $\Omega_\mathrm{lim}=0.63g$.  In (c)
    $\kappa=0.3g$ giving $\kappa_\mathrm{1P}=0.003g$,
    $\kappa_\mathrm{2P}=0.13g$ and $\Omega_\mathrm{lim}=1.15g$.  In
    (a), $\gamma=0.01g$.  In (b), (c), $\gamma=0.1g$.\newline
    \underline{Second row}: Analytical approximations for the
    two contributions to the cavity emission: from single photons,
    $L_1+L_2$, and from pairs of photons,
    $L_\mathrm{I}+L_\mathrm{II}$. Panel (d) corresponds to situation
    (a) and panel (e) to situations (b) and (c), plotted in solid and
    dashed lines, respectively.}
  \label{fig:3}
\end{figure}

The equations under incoherent excitation involve only the
populations:
\begin{subequations}
  \label{eq:TueDec20184307CET2011}
  \begin{align}
    &\partial_t\rho_\mathrm{G,G}=-2P\rho_\mathrm{G,G}+\gamma\rho_\mathrm{V,V}+(\gamma+\kappa_\mathrm{1P})\rho_\mathrm{H,H}+\kappa_\mathrm{2P}\rho_\mathrm{B,B}\,,\\
    &\partial_t\rho_\mathrm{V,V}=P\rho_\mathrm{G,G}-(\gamma+P)\rho_\mathrm{V,V}+\gamma\rho_\mathrm{B,B}\,,\\
    &\partial_t\rho_\mathrm{H,H}=P\rho_\mathrm{G,G}-(\gamma+P+\kappa_\mathrm{1P})\rho_\mathrm{H,H}+(\gamma+\kappa_\mathrm{1P})\rho_\mathrm{B,B}\,,\\
    &\partial_t\rho_\mathrm{B,B}=P\rho_\mathrm{V,V}+P\rho_\mathrm{H,H}-(2\gamma+\kappa_\mathrm{1P}+\kappa_\mathrm{2P})\rho_\mathrm{B,B}\,.
  \end{align}
\end{subequations}
Together with the normalization Tr$(\rho)=1$, they provide analytical
expressions, such as:
\begin{subequations}
  \label{eq:TueDec20185648CET2011}
  \begin{align}
    &\rho_\mathrm{H,H}\approx \frac{P\gamma\Gamma_\mathrm{B}+P^2(\Gamma_\mathrm{B}+\Gamma_\mathrm{H}-\gamma)}{\gamma\Gamma_\mathrm{B}\Gamma_\mathrm{H}+2P\big[\gamma(\Gamma_\mathrm{B}-\Gamma_\mathrm{H}) + \Gamma_\mathrm{B}\Gamma_\mathrm{H}\big]+3P^2\Gamma_\mathrm{B}+2P^3}\xrightarrow{P\rightarrow 0} \frac{1}{\Gamma_\mathrm{H}}P\,, \\
    &\rho_\mathrm{B,B}\approx
    \frac{P^2(\Gamma_\mathrm{H}+\gamma)+2P^3}{\gamma\Gamma_\mathrm{B}\Gamma_\mathrm{H}+2P\big[\gamma(\Gamma_\mathrm{B}-\Gamma_\mathrm{H})
      +
      \Gamma_\mathrm{B}\Gamma_\mathrm{H}\big]+3P^2\Gamma_\mathrm{B}+2P^3
    }\xrightarrow{P\rightarrow
      0}\frac{\Gamma_\mathrm{H}+\gamma}{\gamma\Gamma_\mathrm{B}\Gamma_\mathrm{H}}P^2\,,
  \end{align}
\end{subequations}
where
$\Gamma_\mathrm{B}=2\gamma+\kappa_\mathrm{1P}+\kappa_\mathrm{2P}$ and
$\Gamma_\mathrm{H}=\gamma+\kappa_\mathrm{1P}$ are the dissipation
rates of levels the B and H. At vanishing pumping, we find the
expected linear increase for single exciton populations and square
increase for the biexciton population. In Fig.~\ref{fig:3}(a) we give
an example of the quality of the approximation. Both the numerical
exact solution of the full master equation (solid lines) and the
approximated formulas~(\ref{eq:TueDec20185648CET2011}) (dashed lines)
are plotted for increasing excitation. They match almost perfectly for
the whole pumping range. Eventually the system saturates on the
biexciton state (not shown).

The equations under coherent excitation involve not only the
populations but also some off-diagonal terms of the density matrix:
\begin{subequations}
  \label{eq:TueDec20203043CET2011}
  \begin{align}
    &\partial_t\rho_\mathrm{G,G}=\gamma\rho_\mathrm{V,V}+\Gamma_\mathrm{H}\rho_\mathrm{H,H}+\kappa_\mathrm{2P}\rho_\mathrm{B,B}-i\Omega_\mathrm{V}(\rho_\mathrm{V,G}-\rho_\mathrm{G,V})\,,\\
    &\partial_t\rho_\mathrm{G,V}=\big(i\frac{\chi}{2}-\frac{\gamma}{2}\big)\rho_\mathrm{G,V}+i\Omega_\mathrm{V}(\rho_\mathrm{G,G}-\rho_\mathrm{V,V}+\rho_\mathrm{G,B})\,,\\
    &\partial_t\rho_\mathrm{G,B}=-\big(\frac{\gamma+\Gamma_\mathrm{H}+\kappa_\mathrm{2P}}{2}\big)\rho_\mathrm{G,B}-i\Omega_\mathrm{V}(\rho_\mathrm{V,B}-\rho_\mathrm{G,V})\,,\\
    &\partial_t\rho_\mathrm{V,V}=-\gamma\rho_\mathrm{V,V}+\gamma\rho_\mathrm{B,B}+i\Omega_\mathrm{V}(\rho_\mathrm{V,G}-\rho_\mathrm{G,V}+\rho_\mathrm{V,B}-\rho_\mathrm{B,V})\,,\\
    &\partial_t\rho_\mathrm{V,B}=\big(-i\frac{\chi}{2}-\frac{2\gamma+\Gamma_\mathrm{H}+\kappa_\mathrm{2P}}{2}\big)\rho_\mathrm{V,B}+i\Omega_\mathrm{V}(\rho_\mathrm{V,V}-\rho_\mathrm{B,B}-\rho_\mathrm{G,B})\,,\\
    &\partial_t\rho_\mathrm{H,H}=-\Gamma_\mathrm{H}\rho_\mathrm{H,H}+\Gamma_\mathrm{H}\rho_\mathrm{B,B}\,,\\
    &\partial_t\rho_\mathrm{B,B}=-\Gamma_\mathrm{B}\rho_\mathrm{B,B}+i\Omega_\mathrm{V}(\rho_\mathrm{B,V}-\rho_\mathrm{V,B})\,.
  \end{align}
\end{subequations}
All off-diagonal terms involving the $H$-state vanish in the steady
state. Including the normalization Tr$(\rho)=1$, we obtain analytical
expressions, such as:
\begin{subequations}
  \label{eq:TueDec20203050CET2011}
  \begin{align}
    &\rho_\mathrm{B,B}= \rho_\mathrm{H,H}\approx \frac{4(2\Omega_\mathrm{V}^2)^2}{\Gamma_\mathrm{B}^2\chi^2+\Gamma_\mathrm{B}(\gamma+\Gamma_\mathrm{B})\Omega_\mathrm{V}^2+64\Omega_\mathrm{V}^4}\xrightarrow{\Omega_\mathrm{V}\rightarrow 0}\frac{4(2\Omega_\mathrm{V}^2)^2}{\Gamma_\mathrm{B}^2\chi^2}  \,,\\
    &\rho_\mathrm{V,V}\approx \frac{4\Omega_\mathrm{V}^2(\Gamma_\mathrm{B}^2+4\Omega_\mathrm{V}^2)}{\Gamma_\mathrm{B}^2\chi^2+\Gamma_\mathrm{B}(\gamma+\Gamma_\mathrm{B})\Omega_\mathrm{V}^2+64\Omega_\mathrm{V}^4}\xrightarrow{\Omega_\mathrm{V}\rightarrow 0} \frac{4\Omega_\mathrm{V}^2}{\chi^2}\,,\\
    &\rho_\mathrm{B,G}\approx - i \frac{4\Omega_\mathrm{V}^2\Gamma_\mathrm{B}\chi}{\Gamma_\mathrm{B}^2\chi^2+\Gamma_\mathrm{B}(\gamma+\Gamma_\mathrm{B})\Omega_\mathrm{V}^2+64\Omega_\mathrm{V}^4}\xrightarrow{\Omega_\mathrm{V}\rightarrow 0} -i\frac{4\Omega_\mathrm{V}^2}{\Gamma_\mathrm{B} \chi }\,.
  \end{align}
\end{subequations}
The behaviour at vanishing pumping is the expected one: the exciton
populations increase as $\Omega_\mathrm{V}^2$ and the biexciton
population as $\Omega_\mathrm{V}^4$. Consistently, the two-excitation
off-diagonal term $|\rho_\mathrm{B,G}|^2$ increases as
$\Omega_\mathrm{V}^4$. The population of the biexciton state is
plotted in Fig.~\ref{fig:3}(b) and (c) for two different coupling
strengths with the cavity, strong ($\kappa=g$) and very strong
($\kappa=0.3g$). Again we observe a very good agreement between exact
and approximated solutions. The agreement depends more critically on
$\gamma$ than under incoherent excitation, the larger $\gamma$ the
better the agreement. That is why we increased it from $\gamma=0.01g$
to $\gamma=0.1g$ in Figs.~\ref{fig:3}(b), (c). Of course, this means
slightly decreasing the efficiency of the two cavity photon emission
as compared to the total emission of the
system~\cite{delvalle11d}. However, in this work we are more
interested in identifying what plays a fundamental role in the
dynamics under continuous excitation, in order to grasp the conditions
for two-photon emission. Large pumping leads to saturation which in
this case means equal population for all four dot levels, $1/4$.

Let us next derive analytical expressions for $n_a$ and $Q$, in terms
of the previous analytical matrix elements.

\subsection{Cavity properties}

The spectrum of emission in the steady state reads $\pi S(\omega)=\Re
\int_{0}^{\infty} d\tau e^{i\omega \tau}\lim_{t\rightarrow
  \infty}\langle a^{\dagger} (t) a (t+\tau)\rangle$. At the two-photon
resonance, we can split the spectrum into its four main
contributions:~\cite{delvalle11d}
\begin{equation}
  \label{eq:ThuApr14014048CEST2011}  
  S(\omega)=\frac1\pi\sum_{\alpha\in\{1,2,\mathrm{I},\mathrm{II}\}}\left(L_\alpha\frac{\frac{\gamma_\alpha}{2}}{(\frac{\gamma_\alpha}{2})^2+(\omega-\omega_\alpha)^2}
    - K_\alpha\frac{\omega-\omega_\alpha}{(\frac{\gamma_\alpha}{2})^2+(\omega-\omega_\alpha)^2}\right)
  \,,
\end{equation}
corresponding to the three cavity-mediated transitions in the system,
characterised by their frequency ($\omega_\alpha$) and broadening
($\gamma_\alpha$). $L_\alpha$ quantifies the total intensity emitted
through a given transition~$\alpha=\ket{\mathrm{i}}\rightarrow
\ket{\mathrm{f}}$, from a given initial state $\ket{\mathrm{i}}$ to a
given final state $\ket{\mathrm{f}}$. The sum of all $L_\alpha$ is the
total photon mean number in the steady state:
\begin{equation}
  \label{eq:WedDec14172003CET2011}
  n_a=\text{Tr}(\tilde\rho a^\dagger a)=\sum_{i,f}\tilde \rho_{i,i}|\bra{f}a\ket{i}|^2=L_1+L_2+L_\mathrm{I}+L_\mathrm{II}\,,
\end{equation}
in the basis of states that diagonalises the density matrix. Due to
the dispersive (weak) coupling to the cavity, we can safely assume
that the dynamics never involve more than two photons and truncate the
dot-cavity Hilbert space as in Fig.~\ref{fig:1}(b), (c). Moreover,
only states with no photon are significantly populated.  All other
states remain virtual, in the sense that they serve as intermediate
states for perturbative second order processes but never achieve a
sizable population. As a result, the intitial states that we should
consider are not exactly $\ket{\mathrm{H,0}}$ and
$\ket{\mathrm{B,0}}$, with zero photon, because the dispersive
interaction with the cavity couples each of them weakly to states with
one or two photons. The photonic components can be obtained by
diagonalising the Hamiltonian in each manifold of excitation. For
instance, in the manifold of two excitations, states
$\ket{\mathrm{B,0}}$, $\ket{\mathrm{H,1}}$ and $\ket{\mathrm{G,2}}$
interact through the non-Hermitian Hamiltonian
\begin{equation}
  \label{eq:ThuDec15223352CET2011}
  H_\mathrm{2P}=\left( 
    \begin{array}{ccc}
      -i\gamma & g & 0 \\
      g      & \chi/2-i(\kappa+\gamma)/2   & \sqrt{2}g \\
      0      & \sqrt{2}g   & -i\kappa \\
    \end{array}
  \right)\,,
\end{equation}
as shown in Fig.~\ref{fig:1}(b). We have added the corresponding
dissipation of each level as an imaginary part to the frequency and
considered the renormalization of the coupling by the number of
photons involved ($g$ or $\sqrt{2}g$ for one and two-photon states,
respectively). Diagonalising this Hamiltonian to second order at large
$\chi$ and $\kappa\gg\gamma$, one obtains new eigenstates that differ
from the bare ones in additional small components from all the other
bare states. That is, state $\ket{\mathrm{B}}\approx
\ket{\mathrm{B,0}}$ becomes $\ket{\mathrm{B}_2}\approx C
\ket{\mathrm{B,0}}+
C_\mathrm{1P}\ket{\mathrm{H,1}}+C_\mathrm{2P}\ket{\mathrm{G,2}}$ with
$|C_\mathrm{1P}|=2g/\chi$ and
$|C_\mathrm{2P}|=2\sqrt{2}g^2/(\chi\kappa)$. Similarly, in the
manifold of one excitation, states $\ket{\mathrm{H,0}}$ and
$\ket{\mathrm{G,1}}$ interact through the non-Hermitian Hamiltonian
\begin{equation}
  \label{eq:WedDec21003504CET2011}
  H_\mathrm{1P}=\left( 
    \begin{array}{cc}
      \chi/2-i\gamma/2 & g  \\
      g & -i\kappa/2   
    \end{array}
  \right)\,,
\end{equation}
as shown in Fig.~\ref{fig:1}(c). The state $\ket{\mathrm{H}}\approx
\ket{\mathrm{H,0}}$ becomes $\ket{\mathrm{H}_1}\approx C'
\ket{\mathrm{H,0}}+C_\mathrm{1P}\ket{\mathrm{G,1}}$. In practical
terms, one must consider as initial states in
Eq.~(\ref{eq:WedDec14172003CET2011}) the superpositions
$\ket{\mathrm{B}_2}$ and $\ket{\mathrm{H}_1}$, with coefficient
rewritten as $|C_\mathrm{1P}|^2=\kappa_\mathrm{1P}/\kappa$ and
$|C_\mathrm{2P}|^2=\kappa_\mathrm{2P}/(2\kappa)$.\footnote{In fact,
  this is an alternative way to estimate $\kappa_\mathrm{1P}$,
  $\kappa_\mathrm{2P}$, and then $g_\mathrm{1P}$, $g_\mathrm{2P}$ with
  Eq.~(\ref{eq:MonDec12101215CET2011}). The effective photonic decay
  rate of $\ket{\mathrm{H}_1}$ is its photonic component
  $|C_\mathrm{1P}|^2$ times the associated decay rate $\kappa$,
  etc.} These results do not change within the same degree of
approximation if we include decoherence due to the incoherent pump
(affecting all levels but $\ket{\mathrm{B,0}}$) as long as
$P\ll\chi$. Coherent excitation does not bring any additional
decoherence.

From these perturbed initial states, there are four main possible
transitions via the cavity mode (see Fig.~\ref{fig:1}(a)), at
frequencies and with broadenings that we already
described~\cite{delvalle11d}. We obtain the following analytical
expressions for their intensities:
\begin{itemize}
\item[1)] the decay from $\ket{\mathrm{B}_2}$ to $\ket{\mathrm{H,0}}$,
  gives rise to the component
  $L_1=\rho_\mathrm{B,B}|\bra{\mathrm{H,0}}a\ket{\mathrm{B}_2}|^2=\rho_\mathrm{B,B}\kappa_\mathrm{1P}/\kappa$,
\item[2)] the decay from $\ket{\mathrm{H}_1}$ to $\ket{\mathrm{G,0}}$,
  gives rise to the component
  $L_2=\rho_\mathrm{H,H}|\bra{\mathrm{G,0}}a\ket{\mathrm{H}_1}|^2=\rho_\mathrm{H,H}\kappa_\mathrm{1P}/\kappa$,
\item[I)] the decay from $\ket{\mathrm{B}_2}$ to $\ket{\mathrm{G,1}}$,
  gives rise to the component
  $L_\mathrm{I}=\rho_\mathrm{B,B}|\bra{\mathrm{G,1}}a\ket{\mathrm{B}_2}|^2=\rho_\mathrm{B,B}2
  \kappa_\mathrm{2P}/(2\kappa)$,
\item[II)] the direct decay from $\ket{\mathrm{G,1}}$ to
  $\ket{\mathrm{G,0}}$, gives rise to the component
  $L_\mathrm{II}=\rho_\mathrm{G1,G1}|\bra{\mathrm{G,0}}a\ket{\mathrm{G,1}}|^2=\rho_\mathrm{G1,G1}$. This
  level has a very small population in the steady state. Its dynamics
  under incoherent excitation reads
  $\partial_t\rho_\mathrm{G1,G1}\approx
  \kappa_\mathrm{2P}\rho_\mathrm{B0,B0}-\kappa
  \rho_\mathrm{G1,G1}-2P\rho_\mathrm{G1,G1}$. Then, we have,
  \begin{equation}
    \label{eq:MonDec19123406CET2011}
    \rho_\mathrm{G1,G1}\approx \kappa_\mathrm{2P}/(\kappa+2P)\rho_\mathrm{B,B}\,.
  \end{equation}
  For the success of a simultaneous two-photon state emission from
  $\ket{\mathrm{B,0}}$, the population of $\ket{\mathrm{G,1}}$ should
  remain small, keeping its virtual nature. Moreover, the photon
  should also be quickly emitted before incoherent pumping drives the
  state upwards into $\ket{\mathrm{H,1}}$. This means that we require
  \begin{equation}
    \label{eq:TueDec20140040CET2011}
    P \ll P_\mathrm{lim}=\kappa/2 \quad \text{and} \quad \kappa \gg \kappa_\mathrm{2P}\,.
  \end{equation}
  in order not to break the indistinguishability and simultaneity of
  the two emission events.

  In the case of coherent excitation we set $P\rightarrow 0$. We
  neglect the possible dynamics of state $\rho_\mathrm{G1,G1}$ due to
  the effective one- and two-photon effective driving to the
  off-resonant state $\ket{\mathrm{V,1}}$ and the two-photon resonant
  state $\ket{\mathrm{B,1}}$. They do not play a role as they are
  small in the regime of pumping where our approximations hold:
  $\Omega_\mathrm{1P}\approx \Omega_\mathrm{V}\gamma/\chi$ and
  $\Omega_\mathrm{2P}\approx 2\Omega_\mathrm{V}^2/\chi$.\footnote{One
    can estimate them by comparing populations $\rho_\mathrm{V,V}$ and
    $\rho_\mathrm{B,B}$ in Eq.~(\ref{eq:TueDec20203050CET2011}), to
    second order in $1/\chi$, with the occupation of a two-level
    system in the linear regime, given by~$4\Omega^2/\Gamma$, with
    $\Gamma$ its decay rate.}
\end{itemize}
In total, the cavity intensity reads:
\begin{equation}
  \label{eq:WedDec21113634CET2011}
  n_a\approx \rho_\mathrm{B,B}\big[\frac{\kappa_\mathrm{1P}+\kappa_\mathrm{2P}}{\kappa}+\frac{\kappa_\mathrm{2P}}{\kappa+2P}\big]+\rho_\mathrm{H,H}\frac{\kappa_\mathrm{1P}}{\kappa}\,.
\end{equation}
Similarly, we can obtain the second order coherence function in two
different ways,
\begin{subequations}
  \label{eq:WedDec14172003CET20112}
  \begin{align}
    G^{(2)}=\text{Tr}(\tilde\rho a^\dagger a^\dagger aa)=&\sum_{i,f}\tilde \rho_{i,i}|\bra{f}a^2\ket{i}|^2\approx \rho_\mathrm{B,B} \kappa_\mathrm{2P}/\kappa \\
    &\approx (L_\mathrm{I}+L_\mathrm{II})/2 \approx
    \rho_\mathrm{B,B}\kappa_\mathrm{2P}\big[\frac{\frac{1}{\kappa}+\frac{1}{\kappa+2P}}{2}\big]\,,\label{eqWedDec21160746CET2011}
  \end{align}
\end{subequations}
The two lines converge at low incoherent pumping. 

\subsubsection{Incoherent excitation}

We prefer to use the second line of
Eq.~(\ref{eq:WedDec14172003CET20112}) to compute $G^{(2)}$ in the case
of incoherent pumping because it incorporates the fact that the second
photon in the two-photon de-excitation is less likely to be emitted
due to the pumping induced transition $\ket{\mathrm{G,1}}\rightarrow
\ket{\mathrm{H,1}}$ explained above. We see in Fig.~\ref{fig:3}(a)
that both $n_a$ and $Q$ calculated from
Eq.~(\ref{eqWedDec21160746CET2011}) are in good agreement with the
exact solution. One may think that pumping stronger will be beneficial
for our purposes as the biexciton level is more likely occupied and
available for the two-photon de-excitation. However, when the pumping
overcomes the cavity losses, $P_\mathrm{lim}\approx \kappa/2$, the
$Q$-factor drops dramatically becoming very different from the
analytical result (not shown). The states that we assumed virtual are
no longer so as the pumping forces them to acquire some dynamics. The
cavity intensity increases and a truncation at two photons is not
appropriate.  Even though the coupling is not strong enough to achieve
lasing~\cite{delvalle10a}, the system is not anymore in the
spontaneous emission regime. Finally, the decoherence and disruptive
effect of the pump dominates, quenching also the cavity emission. To
sum up, the two-photon mechanism that we pursue and described
analytically---that is, an efficient succession of fast two-photon
emissions---is washed out when $P>P_\mathrm{lim}$. Remaining in the
unsaturated regime of Fig.~\ref{fig:3}(a), we ensure the survival of
the desired two-photon emission.

The $Q$-factor achieves a maximum at a pumping rate that we call
$P_\mathrm{max}$, marked with a vertical guideline in the figure. At
this point, the pumping is small enough to keep the virtual nature of
$\ket{\mathrm{G,1}}$ and, therefore, the simultaneity and
indistinguishability of the emissions. At the same time, the
population of the biexciton is already clearly larger than the other
states, enough for the two-photon emission to dominate. This is shown
in Fig.~\ref{fig:3}(d) where we compare the intensity associated to
the single photon emissions, $L_1+L_2$, with that associated to the
two-photon emission, $L_\mathrm{I}+L_\mathrm{II}$.  Once the small
pumping that we call $P_\mathrm{min}$ (where both contributions cross)
is overcome, the two-photon emission dominates. Due to the fact that
below $P_\mathrm{min}$ the one-photon emission dominates, the $Q$
factor is zero for vanishing pumping. In the limiting case of
$\gamma=0$, the minimum pumping vanishes and the two-photon process
dominates at all pumpings, depending on the system parameters only,
\begin{equation}
  \label{eq:ds}
  Q_0=\lim_{P\rightarrow 0}( \lim_{\gamma\rightarrow 0} Q)=\frac{1}{3}\frac{4g^2}{4g^2+\kappa^2}\leq \frac{1}{3}\,.
\end{equation}
Note that in this case, $Q$ still increases from $Q_0$ and reaches its
maximum value at a finite $P_\mathrm{max}$, as when $\gamma\neq 0$.

\subsubsection{Coherent excitation}

In the case of coherent excitation, we also find a maximum value of
excitation intensity, $\Omega_\mathrm{lim}$, after which our
analytical expressions do not hold, as shown in the two different
examples of Fig.~\ref{fig:3}(b), (c). In principle, coherent
excitation does not induce decoherence at high pumpings. However,
passed the linear regime, it starts dressing the quantum dot
four-level system, shifting the levels and changing its
resonances. This becomes prejudicial as well for our two-photon
emission process. One would need to recalculate the conditions for a
two-photon resonance taking into account the dressing by the laser. A
more careful analysis of the spectrum of emission (and the new peaks
appearing) would then be required. Moreover, the two-photon absorption
from the laser creates coherence between states $\ket{\mathrm{G}}$ and
$\ket{\mathrm{B}}$ that may interfere with our mechanism. We already
showed an example of such laser-cavity interaction in
Fig.~\ref{fig:2}(i) and (ii). We can estimate $\Omega_\mathrm{lim}$ as
the point at which $\Omega_\mathrm{2P}=\kappa_\mathrm{2P}$, and get
$\Omega_\mathrm{lim}\approx 2\sqrt{2}g^2/\sqrt{\chi\kappa}$.
Similarly to the incoherent pumping case,
$\Omega_\mathrm{V}>\Omega_\mathrm{lim}$ also leads to a growth in the
cavity emission that we cannot reproduce analytically due to the
truncation of the Hilbert space at two-photons.

In contrast with the incoherent excitation, the $Q$-factor starts from
a local maximum at vanishing pumping. The reason is that the biexciton
state has always the same population as the H-state so there can be a
large ratio of two versus one-photon emission for arbitrarily small
pumping, depending on $\kappa_\mathrm{2P}/\kappa_\mathrm{1P}$ only
(always $\gg 1$ in our examples). This is shown in Fig.~\ref{fig:3}(e)
where two-photon process always dominates,
$L_\mathrm{I}+L_\mathrm{II}>L_1+L_2$. The fraction
$L_\mathrm{I}+L_\mathrm{II}/(L_1+L_2)=4g^2/\kappa^2$ is indeed
constant over the whole region. Then, we can conclude that the maximum
$Q$-factor achieved in the region of interest (before saturation)
reads
\begin{equation}
  \label{eq:WedDec21200516CET2011}
  Q_0=Q_\mathrm{max}=\lim_{\Omega_\mathrm{V}\rightarrow 0}=\frac{1}{2}\frac{4g^2}{4g^2+\kappa^2}\leq \frac{1}{2}\,.
\end{equation}
This puts a upper limit to $Q$ of~$1/2$ in the present conditions. It
also tells us that the better the system (deeper into the strong
coupling regime), the higher the bunching in the linear regime,
c.~f. Fig.~\ref{fig:3}(b) and (c).  However, we must bear in mind that
the state $\ket{\mathrm{G,1}}$ should remain virtual in order to have
simultaneous and indistinguishable emissions, and this means keeping
$\kappa_\mathrm{2P}/\kappa\ll 1$, that is, $\kappa \gg 4g^2/\chi$
(equal to $0.2g$ in our examples).

\section{Conclusions}
\label{sec:3}

Let us put together the conditions to achieve an efficient emission of
two simultaneous and indistinguishable photons through the cavity mode
under continuous excitation:
\begin{equation}
  \label{eq:ThuDec22000820CET2011}
  \frac{4g^2}{\chi}< \kappa < 2g\quad \text{and} \quad P_\mathrm{min}<P< \kappa/2 \quad \text{or} \quad \Omega_\mathrm{V}<\frac{4g^2}{\sqrt{2\chi\kappa}}\,.
\end{equation}
The condition $\kappa < 2g$ ensures high probability of the two-photon
the spontaneous emission, as in the case of a decay from the initial
state $\ket{\mathrm{B,0}}$~\cite{delvalle11d}. The condition $\kappa
>4g^2/\chi$ ensures a fast photon pair emission, avoiding accumulation
of photons in the cavity, that may break simultaneity or
indistinguishability. This is the same as for a single photon emitter:
it is desirable that the system remains in the weak coupling or
Purcell regime so that quantum properties are preserved. The
conditions on the excitation rates go also in this direction, to avoid
saturation of any kind. But excitation must, at the same time, be
enough to populate noticeably the biexciton state. In this sense,
although $P_\mathrm{min}$ is generally very small, the V- polarised
laser excitation of the system, succeeds to probe the system at an
arbitrarily weak driving, giving similar results as looking into the
spontaneous emission directly.

\acknowledgments     
 
We acknowledge support from the Alexander von Humboldt Foundation, the
FPU program (AP2008-00101) from the Spanish Ministry of Education and
the FP7-PEOPLE-2009-IEF project SQOD.


\bibliography{Sci,2phot}   
\bibliographystyle{spiebib}   

\end{document}